# Finding optimal hull shapes for fast vertical penetration into water

**Pingan Liu, Zhihao Zhang, Lei Zheng**

College of Aerospace and Civil Engineering,
Harbin Engineering University,
Harbin 150001, China

**Igor Nesteruk**[1]
Institute of Hydromechanics,
National Academy of Sciences,
03057 Kyiv, Ukraine
e-mail: inesteruk@yahoo.com

**ABSTRACT**

*A new approach for the supercavitating hull optimization was proposed, which combines the CFD simulation and analytical methods. The high-speed penetration into water at the velocity 1 km/s was considered with the fixed body mass and caliber. Six different axisymmetric hulls with disc, 30° and 10° conical cavitators were simulated with the use of FLUENT at the first stage of penetration (till the 6 m depth). The optimized hull shapes were calculated with the use of quasi-steady approach and the characteristics obtained by CFD simulation. The use of slender cavitators drastically decreases the local pressures, but the higher friction drag also reduces the depth of supercavitating penetration. Optimized conical hulls can ensure much deeper penetration in comparison with the initial cylindrical hulls. The slenderest cavitator with the cylindrical hull ensures small loads and very deep penetration.*

[1] Corresponding author. e-mail: inesteruk@yahoo.com

## INTRODUCTION

The high speed penetration into water is a very complicated phenomenon, connected with the creation and development of the non-steady cavity and high loads acting on the body. There are not a lot of investigations focusing on water entry of bodies of revolution at the high speed about 1 km/s. Some experimental results are presented in [1-3]. McMillen [4] engaged in experimental research on high-speed sphere entry, and captured shock wave propagation in water. Lundstrom and Fung [5] studied water entry at speed of 1km/s experimentally, and predicted the radius of cavity. M. Lee et al. [6, 7] used the results of [5] to analyze the sub- and supersonic water entry theoretically and numerically. Meanwhile some experimental results on supercavitating projectile (truncated cone) was reported, which are in good agreement with numerical results, [8]. Hrubes [9] captured the shape of cavity and shock wave by high speed imaging systems at the moving projectile with the sub- and supersonic velocities. Neaves and Edwards [10, 11] used a low-diffusion method to calculate the flow field parameters, fitting well with the experimental results of Hrubes. The influence of different projectile nose shapes on the water entry was studied by a series of numerical simulations using the AUTODYN-2D [12]. Sun et al. [13] conducted research on a supersonic body of revolution with the use of CFD simulation, and the instantaneous drag coefficient of the body contacting the water surface was reduced. Multi-material Arbitrary Lagrange Euler (ALE) methods were used to simulate the process of projectile penetration at speed range 300-1500m/s in [14]. It was shown that the speed attenuation coefficient and the drag coefficient increase with the increase of the impact velocity. Rashidi et al. [15] simulated the supercavitating flow with the use of Fluent to optimize the cavitator shape.

If we need to solve a problem of optimization, for example, to increase of depth of high-speed immersion (without the loss of supercavitation flow pattern) or to reduce the loads arising in the body, the capacity of CFD methods is very limited. It is connected with large sizes of calculating domains and high time costs for the optimal body shape searching.

Analytical methods developed in [16-20] to increase the range of the supercavitation motion, allow calculating the optimal body shape for different isoperimetric conditions. This quasi-steady approach uses the semi-empirical Garabedian formulas [21] for the cavity shape, but their accuracy in compressible fluid raises questions. It is also not applicable for the initial stage of the penetration, when the flow is very unsteady.

In this paper we combine CFD and analytical solutions to obtain an effective optimization tool to diminish loads and to increase the range of penetration without loosing high penetration speed (which must be large enough to preserve the supercavitating flow pattern). To calculate the flow characteristics for the initial stage of penetration (till the depth of 6 m), we will use Fluent software. The calculated drag and cavity pressure values will be used in quasi-steady approach to find the optimal shape of the body in order to increase the depth of high-speed penetration and to diminish the loads acting on the body.

## 2 RESEARCH METHOD

### 2.1 Main Characteristics of the Projectile and Fluids

Since the solution of any optimization problem depends on the isoperimetric conditions, we fix the maximal body diameter $D_b$ = 0.125 m, its mass $m$ = 40 kg and the initial velocity of penetration $U_0$ = 1000 m/s. We will consider only vertical penetration into water. For calculation we will use the values of the atmospheric pressure $p_a$ = 101.325 kPa; the speed of sound in water $a_0$ = 1484.6 m/ at atmospheric pressure; ambient temperature of 300 K; the kinematic viscosities of water $v$ = 1.01E-06 $m^2$ /s and gas $v_g$= 14.8E-06 $m^2$ /s. The reference density of air $\rho_g$ is 1.225 kg/$m^3$, and saturated vapor pressure $p_v$ is 3540 Pa. Thus, the body movement in air is supersonic and subsonic in water with the initial Mach number, $M = U_0/a_0 \approx 0.674$.

Two compressible fluid phases were used in calculations. The gas is assumed to be ideal. The compressibility of water can be described by Murnaghan-Tait equation of state [22], ignoring the influence of temperature:

$$p = \frac{K_0}{n}[(\frac{\rho}{\rho_0})^n - 1] + p_a \tag{1}$$

Here $\rho_0$ = 998.2 kg/$m^3$ is the reference density of water, $K_0$ = 2.2E+09 Pa is the reference bulk modulus, $n$ = 7.15 is the density index. The speed of sound in water is calculated by Newton-Laplace equation [23]:

$$a = \sqrt{\frac{K_0 + n(p - p_a)}{\rho}} \quad (2)$$

Neglecting the tail fins we may simplify the projectile to a 2D body of revolution. Since the length of the projectile is not fixed we will use 3 initial hull shapes with the cylinders of length 0.6 m and diameter of 0.125 m. In order to estimate the load decrease, three different cavitators (disc, 30° and 10° cones) were used. The maximum diameter of all the cavitators is 0.125 m and coincides with the diameter of the cylindrical part of the hull, see. Fig. 1 (1a, 2a, 3a). These initial shapes will be improved to increase the depth of high-speed penetration, see Fig. 1 (1b, 2b, 3b).

**2.2 Description of CFD Method**

In this paper we used the commercial CFD Fluent software in 2D axisymmetric space. If the fins on the hull are neglected and the penetration is vertical, the flow pattern is expected to be axisymmetric. Therefore, the selected 2D axisymmetric flow pattern can be used in our case. The VOF (volume of fluid) model (including two phases: water and gas) [24] is selected to calculate the free boundary. Then the density and viscosity of fluid can be expressed by liquid volume fraction [25]. We use also the Schnerr-Sauer cavitation model [26] based on the Rayleigh-Plesset bubble kinetic equation [27] to simulate the cavitation inception and development. The water and gas flow are supposed to be turbulent (we use the realizable *k*-*ε* turbulent model [28]). All these models are embedded in the Fluent software.

The six DOF (six degree of freedom) solver computed external forces and moments on projectile [29]. Gravity force is taken into account in the Fluent software, but its influence on the body dynamics is minimal, since the Froude numbers are very high. The solution model - PISO (pressure implicit with splitting of operators) [30]- is used for calculation of Mass, Momentum and Energy conservation equations. The pressure field was discretized using PRESTO (pressure staggering option) scheme [31]. Modified HRIC (modified version of the high resolution interface capturing) method [32] was used for the volume fraction. The second order upwind scheme was used for the discretization of other physical quantities, like density and momentum. The use of double precision scheme enables to improve the accuracy of calculations. The time step was set to be 5E-07 s, which gives 20 times steps per cycle to ensure the minimum grid height for each advance.

**2.3 Testing of the Code, Mesh and Domain independence**

The computed area is 6m wide, the air height is 1.9m, and the water depth is 10m. The boundary conditions are shown in Fig. 2. Both upper and lower sides are pressure outlet, the right is a wall, and the left is the axis. The rear wall of the projectile is set as a pressure inlet, specifically for making the calculation easy to converge. The pressure on the body base is set as atmospheric one, while the pressure in water increases with the increase of water depth.

To verify the reliability of the calculation model, the truncated cone (10 cm in length, 8 mm in diameter and with a disc cavitator of 4 mm, tested by Schaffar [8]) was selected. The comparison of the experimental and calculated cavity shapes is shown in Fig. 3, where *h* is the water depth, and *R* is cavity radius (the horizontal line is the axis of symmetry). In can be seen that there is a little difference between the results of CFD simulations and measurements.

The initial body with disc cavitator (1a) is selected for testing the computational method. Then, the influence of the mesh and the calculation domain were considered independently. First, four sets of grids were tested: grid 1, grid 2, grid 3 and grid 4. In particular, the height of first grid layer from the body surface was 0.5mm, 0.4mm, 0.25mm and 0.125mm, respectively. The changes of the drag coefficient $C_x = 2X/(\pi\rho_0 U^2 R_n^2)$ and velocity drops are shown in Fig. 4 for different meshes. Here, *X* is the total drag on the entire hull; *U* is the instantaneous velocity of body; $R_n$ is the radius of the cavitator. We will use as well pressure and friction drags on the cavitator to calculate corresponding drag coefficients $C_{xp}$, $C_{xf}$ and the ratio $\gamma = C_x / C_{xp}$. There are some subtle differences in the $C_x$ curves, while the velocity attenuations are basically the same. Hence, the 0.4mm height of first layer of the grid is selected to apply in the following simulations.

Then three different domains were used to consider the effect of their dimensions. The widths of the domains were: 2m, 4m and 5m respectively with the constant water depth 10m. Fig. 5 represents the pressure contours for different domains. Our calculations show that for the domain width greater than 6 m the pressure contours do not change significantly. Therefore, the 6m domain width is selected for the following simulations.

**2.4 Quasi-steady Calculations of the Optimal Hull Shape and the Maximal Range of Supercavitating Motion on Inertia**

The supercavitating inertial motion can be supposed to be quasi-steady when the condition:

$$Q_s = \frac{\pi\rho C_x R_n^2 L_b}{2m} << 1 \tag{3}$$

is satisfied, [19]. Here $L_b$ is total length of the hull, including the length of the cavitators and the cylindrical or conical part of the hull. We will show in Section 3 that inequality Eq. (3) holds after immersion at the depth greater than 6 m for all the cases we calculated.

If we neglect changes of the drag coefficient, then the distance passed by the body is defined by the following formula, (see, e.g., [16]):

$$S = -\frac{2m}{\rho C_x \pi R_n^2} \ln \bar{U} \tag{4}$$

Where $\bar{U}$ is the dimensionless final velocity based on the initial one $U_i$. If the cavitator radius, the final cavitation number $\sigma_f$, the pressure in the cavity $p_c$ and $U_i$ are known, then the depth $h_f$, which can be achieved without loosing the supercavitating flow pattern (before the cavity touches the hull), can be obtained by solving Eq. (4) and the equation

$$\sigma_f = \frac{2(p_a + \rho g h_f - p_c)}{\rho U_i^2 \bar{U}^2} \tag{5}$$

since for the vertical penetration, $h_f$ coincides with $S$ or differs by some known value of initial depth $h_i$ (in particular, we will use $h_i$ = 6 m for some calculations).

It was shown in [16-20] that it is possible to find the optimal cavitator radius and the optimal hull shape, which ensure the maximum range $S^*$. In particular, for disc or non-slender conic cavitators (with the angle $2\theta$, $\theta$ >25°), the semi-empiric Garabedian formulas, [21]:

$$R^2 = \frac{x(1-x)}{\lambda^2}, \quad \frac{R_n}{L} = \frac{\sigma}{2\sqrt{-C_{xp}\ln\sigma}} \tag{6}$$

$$\lambda = \frac{L}{D} = \sqrt{\frac{-\ln\sigma}{\sigma}}, \quad \frac{D}{R_n} = 2\sqrt{\frac{C_{xp}}{\sigma}} \tag{7}$$

can be used to estimate the closed cavity shapes at different moments of time. Where $R$ is the cavity radius, based on the cavity length; $\lambda$ is the cavity aspect ratio; $D$ is the maximal cavity diameter; $L$ is the cavity length; $\sigma$ is the time dependent cavitation number at the cavitator immersion depth. If follows from Eq. (7), that

$$4C_{xp}R_n^2 / D^2 = \sigma \tag{8}$$

Eq. (8) means that the drag coefficient based on the maximal cavity diameter, $4C_{xp}R_n^2 / D^2$ is equal to the cavitation number (see, e.g., Logvinovich's book [33]). Eq. (8) can be also applied for slender cavitators at small enough cavitation numbers. In particular, it was shown in [16, 19] that corrections are approximately 3-7% in for conic cavitators with angles $2\theta$ = 10° and $2\theta$ = 30° at $\sigma \backsim 0.01$.

Equations (4), (8) allows calculating the optimal values of final velocity $\bar{U}^*$, range $S^*$, and final cavitation number $\sigma^*$ ( see [16]):

$$\bar{U}^* = e^{-0.5} \approx 0.607 \tag{9}$$

$$\bar{S}^* = \frac{S^* D_b}{U_0}\sqrt{\frac{\rho g}{m}} = 0.5\left[-\bar{h}_1 + \sqrt{\bar{h}_1^2 + 8/(e\pi)}\right] \tag{10}$$

$$\bar{\sigma}^* = \frac{\sigma^* D U_0}{2}\sqrt{\frac{\rho}{gm}} = (\bar{h}_1 + \bar{S}^*)/(\bar{U}^*)^2 \tag{11}$$

Where $\bar{h}_1 = (h_i + p_a - p_c)D\sqrt{(\rho g/m)}/U_0$. Pressures $p_c$ and $p_a$ must be measured in meters of the water column. Knowing $\sigma^*$ and the given value of the maximum body radius, which at the final moment of the supercavitating inertial motion must be equal to the maximum cavity diameter, the optimal cavitator radius can be calculated with the use of Eq. (8).

The CFD methods of this study will give an opportunity to calculate the pressure drag and to re-estimate $C_{xp}$. Unfortunately, a re-estimation of $D^2/R_n^2$ with the use of CFD is connected with significant problems due to the increase in the sizes of meshes. Thus, we cannot calculate changes in Eq. (8) corresponding to a non-zero Mach number $M$:

$$\frac{4C_{xp}(M)R_n^2}{D^2(0) + \Delta D^2(M)} = \sigma\alpha_1(M) \tag{12}$$

But for slender cavitators and cavities it is possible with the use of the slender body theory developed in [34-39]. In particular, it was shown in [36] that the increase in the cavity radius, connected with the water compressibility can be estimated as follows:

$$\Delta R^2(M) = -\frac{x^2\sigma\ln\omega}{2\ln^2\beta} \tag{13}$$

where

$$\omega = \sqrt{\left|M^2 - 1\right|}, \quad \beta = tg(\theta)$$

and $\theta$ is the half angle of the conical cavitator. The coordinate $x_{max}$ corresponding to the cavity cross-section of the maximal thickness and its maximal diameter can be estimated by differentiation of the first approximation equation, [38]

$$\frac{R^2}{R_n^2} = \alpha\frac{x^2}{R_n^2} + 2\beta\frac{x}{R_n} + 1, \qquad \alpha = \frac{\sigma}{2\ln\beta} \tag{14}$$

Thus,

$$x_{\max} / R_n = -\beta / \alpha \approx 0.5L / R_n \tag{15}$$

$$D^2 / R_n^2 = 4(1 - \beta^2 / \alpha) \tag{16}$$

Putting relationship (15) into Eq. (13) we can obtain

$$\frac{\Delta D^2}{R_n^2} = -\frac{2\beta^2 \ln\omega}{\sigma} \tag{17}$$

The corresponding increase in pressure drag on slender cavitators can be easily estimated for our case of small cavitation numbers, when $C_{xp} \approx C_{x0}$ and

$$C_{x0}(M) = -2\beta^2\left(\ln\frac{\beta\omega}{2} + 1\right) \tag{18}$$

for the subsonic movement in water (see [36, 39]).

Putting Eq. (17) and Eq. (18) into Eq. (12) and taken into account that $D^2(0)/R_n^2 = 4C_{xp}(0)/\sigma$ (according to Eq. (8)), we can obtain the correction function:

$$\alpha_1(M) = \frac{\ln(0.5\beta\omega) + 1}{\ln(0.5\beta) + 1 + 0.25\ln\omega} \tag{19}$$

The results of calculations with the use of Eq. (19) for three different conical cavitators are presented in Fig. 6. It can be seen that the correction function increases with the increasing the Mach number and the cone angle. (E.g., the value of $\alpha_1$ is approximately 1.23 at $M$ = 0.7 for the 30° cone). It means that this correction must be taken into account by dividing the optimal range by the value of $\alpha_1$ (according to Eq. (4) and Eq. (12)). For disc cavitator, the corresponding compressibility correction is unknown. We will use the value for 30° in our calculations of the maximal range, but probably it can yield too optimistic estimation for the depth of the high-speed penetration for the hull 1b shown in Fig 1.

Slender cavitators are preferable to decrease the compressibility correction coefficient $\alpha_1$, but the friction drag on there surface can considerably increase the total drag. To estimate the corresponding increase in the drag (and the reduction of the range), the coefficients

$$\alpha_L = \frac{\pi C_{x0} R_n^2 + C_V V_c^{2/3}}{\pi C_{x0} R_n^2} \tag{20}$$

$$\alpha_T = \frac{\pi C_{x0} R_n^2 + C_T S_c}{\pi C_{x0} R_n^2} \tag{21}$$

were introduced for the laminar and the turbulent flow respectively. Here $V_c$ and $S_c$ are the volume of cavitator and its surface wetted by water; the friction drag coefficient for the laminar unseparated flow, [40]:

$$C_V = \frac{4.7}{\sqrt{\mathrm{Re}_V}}, \quad \mathrm{Re}_V = \frac{U V_c^{1/3}}{\nu} \tag{22}$$

and for the turbulent one, [41]:

$$C_T = \frac{0.0307}{\mathrm{Re}_L^{1/7}}, \quad \mathrm{Re}_L = \frac{U L_c}{\nu} \tag{23}$$

## 3 RESULTS AND DISCUSSION

### 3.1 Calculations of the Optimized Hull Shapes in the First Approximation

We neglect the non-steady processes, which occur at first stages of penetration, and use Eq. (8)-(11) to calculate the first approximation of the optimal cavitator radius. The values of the parameters are $h_i$ = 0 m; $U_i$ = 1000 m/s; $p_a$ =10 m; $p_c$ = 0 m. The calculated value of the optimal final cavity number $\sigma^*$ = 0.0134. The optimal cavitator radius were calculated with the use of Eq. (8), $C_{xp}$ = 0.82 for disc and Eq. (18) for conical cavitators. The results are shown in Table 1 (upper line 1). The optimal length of the conical part of the hulls was calculated with the use of Garabedian formulas (6), (7) for disc and 30° cone cavitator. For slenderer 10° conical cavitator Eq. (15) was used. The results are shown in upper line 3 (In the future we will avoid repetition "Table 1" and use only the line number). Corresponding shapes of the optimal hulls are shown in Fig. 1 (1b, 2b, 3b).

In Table 1 we show also other geometrical characteristics of the hulls: the length of the cavitator $L_c$ (line 2); the length of the conical part of the hull $L_1$ (line 3); the length of the cylindrical part of the hull $L_2$ (line 4); total volume of the hull $V$ (including cavitator) and the volume of cavitator $V_c$ (lines 5); the surface area of the conical cavitator $S_c$ (line 6).

### 3.2 CFD Results. Comparison of the Characteristics of the Initial and Optimized Hulls

For all six body shapes shown in Fig. 1, the CFD simulations have been performed. The results are presented in Figs. 7 and 8 and in the Tables 1 and 2. The body speeds are almost constant in air. During touch and impact the water surface, the speed drops as shown in Fig. 7(a). The speed decrease is very rapid for the initial shapes 1a and 2a in comparison with the hulls 3a and 3b (with the slenderer 10° cone cavitator). The maximal local pressures are much higher on the disc cavitators (bodies 1a and 1b) in comparison with the 30° conical cavitator (bodies 2a and 2b) and, especially, with the 10° one (bodies 3a and 3b, see line 19). For initial shapes 1a, 2a and 3a, the speed attenuates much quickly in comparison with the optimized shapes 1b, 2b and 3b (see Fig. 7b, line 7). In particularly, the speed of body 1a (with the disc cavitator) decreases more than half at the depth of 6 m (the values of speed $U_6$ at the depth $h$ = 6 are presented in line 7).

The total drag coefficients $C_x$ versus depth $h$ (the body touches the water at $h = - L_c$ ) are shown in Fig. 8. The very high peak values of the total drag occur on disc cavitators. For conical cavitators the function $C_x$ ($h$) increases much more smoothly. We can see also, that after attaining some maximum, this function is almost constant for all the hulls. This fact allows applying the assumption about the constant value of $C_x$, used in the theoretical approach [16-20] and formulas (4), (9)-(11). Small enough value of the coefficient $Q_s$ (see (3) and line 13) allows using the assumption about the quasi-steady nature of the body movement.

The results of the CFD drag calculations are presented in lines 10-12. The values of the drag coefficients are based on the velocity $U_6$ and illustrate the that almost all the drag forces are concentrated on the cavitator $C_x \approx C_{xp} + C_{xf}$ . Since the drag coefficient $C_x$ can be treated as a measure of loads acting on the hull, we can conclude that the use of the 10° cavitator (instead of disc) can diminish the average loads approximately 16 times (see line 12). The local pressures on this cavitator are 75 times less in comparison with the disc (see line 19). Slender cavitators yield also smaller values of compressibility correction coefficient $\alpha_1$ (see line 15).

Nevertheless, the use of slender cavitators can be limited by high levels of friction drag and coefficient $\gamma$ (see lines 10 and 14). E.g., the maximal depth of supercavitating inertial motion must be obtained from Eqs. (9)-(11) by dividing the value of range by coefficient $\gamma\alpha_1$. Since for the first approximation of the optimized hulls 1b, 2b and 3b, the value of the range is the same, the values of depth are smaller for conical cavitators (see upper line 18).

The range of supercavitating inertial motion can be increased with the use of changeable shape of the cavitator [18]. Therefore, for the initial stages of penetration a slender cavitators can be used to diminish the loads. After reaching some depth (e.g. 6 m), the shape of cavitator can changed to diminish the area wetted by water and friction drag. With the use of this technology it is possible to have both small pressures on the cavitator and a large penetration depth.

We have done also some theoretical estimations for the drag and coefficients $\alpha_L$ and $\alpha_T$. For this purpose we used the speed $U_6$ (line 7), the Reynolds numbers based on the total hull volume $\mathrm{Re}_V$ and the length of the cavitator $\mathrm{Re}_L$ (see lines 9) and Eqs. (20)-(23). The results are presented in lines 10, 16 and 17. It can be seen that CFD results for friction drag are rather close to the theoretical turbulent estimations and much higher than the laminar ones. It is not surprising, because the high values of the Re numbers do not allow the boundary layer to be purely laminar on the entire surface of the cavitator.

Formula (18) yields a very good estimation for the pressure drag on conical cavitators. E.g., the difference between theoretical and CFD results does not exceed 8-12% (see line 11). The pressure drag coefficients for disc are slightly higher than the theoretical value 0.82 in incompressible water (line 11).

### 3.3 Second Approximation for the Optimized Shapes

After achieving the depth of 6 m, the flow pattern can be treated as quasi-steady with new calculated values of the pressure drag coefficients shown in upper line 11. Then we can use Eqs. (8)-(11) to calculate the second approximation of the optimized shapes, but to improve the accuracy, we need information about the pressure in the cavity, which is not constant. In particular, Abelson [1] used to measure pressure in the water-entry cavity. He found that cavity pressure decreases with the increase of entry speed and is close to zero at 163 m/s. Some theoretical results about the pressure in the ventilated cavities can be found in [42-44].

In Table 2 we show pressures at three selected points in cavity at the last moment of CFD simulation (at depth 6 m). The positions of points are shown in Fig. 9. The pressures vary for different hulls, but for cone cavitators (bodies 2a, 2b, 3a and 3b), the pressure near side wall of the body (points 1 and 2) is close to the saturated vapor pressure. The pressures after body (point 3) are higher but far less than the atmospheric one.

We used Eqs. (8)-(11) with values of parameters $h_i$ = 6 m, $U_i$ =$U_6$ (from line 7) $p_a$ = 10.36 m (calculated for selected values of $p_a$ in Pa and for the reference density of water). We can use the averaged values of pressure at points 1 and 2 (see Table 2) to estimate the pressure in the cavity at the final moment of time. Then $p_c$ = 1.08 m for optimal body 1b and $p_c$ =0.36 m for optimal hulls 2b and 3b. In any case the pressure in the cavity is much smaller than atmospheric one and its influence to the optimal hull shape is not expected to be significant.

Neglecting the changes in the ambient water density (putting in Eqs. (9)-(11) $\rho = \rho_0$) we can calculate the optimal values of the cavitation numbers and then the optimal cavitator radius (with the use of the calculated by CFD values of the pressure drag coefficients listed in upper line 11 and Eq. (8)). The results are shown in lower line 1. The optimal length of the conical part of the hulls was calculated with the use of Garabedian formulas (6), (7) for disc and 30° cone cavitator. For slenderer 10° conical cavitator Eq. (15) was used. The results are shown in lower line 3.

It can be seen that the second approximation of the optimized shapes is very close to the first one and allows increasing the depth of the supercavitating penetration $h_f$ shown in the lower line 18. The accuracy of Eq. (14) is rather limited. To improve the results, the second approximation for the cavity shape in compressible fluid (see [33, 34]) can be used.

To calculate the volumetric total drag coefficients $C_V$, the values from lines 9 and 12 were used. The results (line 20) show that the values of $C_V$ are much lower for optimized hulls. The drag on optimized supercavitating hulls is also lower in comparison with the unseparated shapes of the same volume, since their $C_V$ values vary from 0.01 to 0.015 at high volumetric Reynolds numbers [45].

### 3.4 Depth Calculations for the Initial Cylindrical Hulls

For the cylindrical hulls 1a, 2a and 3a the final moment of the supercavitating inertial motion corresponds to the touching the cavity by their bases. Therefore we cannot use the condition $D_b = D$ and Eqs. (8)-(11). For the disc and the 30$^o$ conical cavitator we can use the condition $L_2 = L$ and Garabedian formulae (6), (7). For slenderer 10° cone the condition $L_2 = 2x_{\max}$ and Eqs. (14)-(16) can be used. These conditions allow calculating the final cavitation numbers. Then the final speed and the depth of supercavitating penetration can obtained with the use of Eqs. (4), (5) and the values of total drag coefficient (calculated in CFD simulation). The results are presented in lines 8 and 18.

For bodies 1a, 1b, 2a, 2b, the comparison of the high-speed penetration depth $h_f$ show that the use of optimized conical hulls is preferable. In the case of the slenderest cavitator (bodies 3a and 3b), the depth can be larger for the cylindrical hulls. This fact can be explained by small differences in the initial velocity $U_6$ (see line 7) and much lower value of the final velocity for the cylindrical hull 3a (see line 8). Since the values of $h_f$ are very close for bodies 3a and 1b, and the average and peak loads on the slender cavitation are much smaller (see lines 11 and 19), the use of cylindrical hulls with slender conical cavitators looks very promising.

To determine the optimal length of a cylindrical hull with 10° conical cavitator, we will do a simple analysis. With the growth of the length $L_2$, the final cavitation number is reduced (see Eq. (15)), and the final dimensionless speed $\overline{U}$ increases (see Eq. (5)). Then according to Eq. (4) depth decreases. Therefore, the optimal are hulls with the minimum length of the cylindrical part. On the other hand, the value of the length limits the accuracy of the used slender body theory (which is applicable for small ratios $D_b/L_2$). Therefore, for the body of the specified maximum diameter $D_b$= 125 mm, the length $L_2$= 600 mm can be recommended as close to optimal, since the ratio $D_b/L_2$ is still not very high. For the final selection of the optimal length $L_2$, it is necessary to use the second approximation (instead of Eq. (14), see [36, 37]) to improve the accuracy of calculations.

**4 CONCLUSIONS**

A new approach for the supercavitating hull optimization was proposed, which combines the CFD simulation and analytical calculations. The high-speed penetration into water at the velocity 1 km/s was considered, the body mass and its maximal diameter were fixed by values 40 kg and 125 mm respectively. In order to investigate the loads and the local impact pressures, six different axisymmetric hulls with disc, 30° and 10° conical cavitators were simulated with the use of FLUENT at the first stage of penetration (till the 6 m depth). The optimized hull shapes were calculated with the use of quasi-steady approach and the characteristics obtained by CFD simulation.

The use of slender cavitators drastically decreases the local pressures (e.g., they are 75 times less on 10° cone in comparison with the disc), but the higher friction drag on slender cavitators reduces the depth of supercavitating penetration (e.g., they are 1.5 times less for the optimized hull 3b with 10° cone cavitator in comparison with the disc one 1b). Optimized conical hulls can ensure much deeper penetration (e.g., 203 m for body 1b) in comparison with the initial cylindrical hulls (e.g., 23 m for body 1a). The slenderest cavitator with the cylindrical hull (body 3a) ensure small loads and very deep penetration (198 m).

Both CFD simulation and quasi-steady calculations need to be improved to increase the accuracy of predictions and the range of their applicability. For example a 3D CFD simulation could be able to simulate non-vertical penetration and to investigate the stability of body motion. It is possible to calculate the optimal shapes for supercavitating inertial motion at any angle to the horizon with the use of the quasi-steady approach, but the accuracy of determining the cavity dimensions in compressible water has to be improved.

**NOMENCLATURE**

| | |
|---|---|
| $a$ | speed of sound |
| $a_0$ | initial speed of sound |
| $C_T$ | drag coefficient for turbulent flow |
| $C_V$ | volumetric drag coefficient |
| $C_x$ | total drag coefficient |
| $C_{x0}$ | pressure drag coefficient at zero cavitation number |
| $C_{xp}$ | pressure drag coefficient on the cavitator |
| $C_{xf}$ | friction drag coefficient on the cavitator |
| $D$ | maximal diameter of the cavity |
| $D_b$ | maximal body diameter |
| $g$ | acceleration of gravity |
| $h$ | depth of movement |
| $h_i$ | initial depth of movement |
| $h_f$ | final depth of movement |
| $\bar{h}_1$ | parameters determined by Eq. (11) |
| $K_0$ | reference bulk modulus (at atmospheric pressure, $p_a$) |
| $L$ | length of the cavity |
| $L_b$ | total length of the hull |
| $L_c$ | length of the conical cavitator |
| $L_1$ | length of the conical part of the hull |
| $L_2$ | length of the cylindrical part of the hull |
| $M$ | Mach number |
| $m$ | body mass |
| $n$ | density exponent |
| $Q_s$ | coefficient, Eq. (3) |
| $p$ | pressure |
| $p_a$ | atmospheric pressure |
| $p_c$ | pressure in the cavity |
| $p_v$ | saturated vapor pressure |
| $R$ | radius of the cavity |
| $\mathrm{Re}_V$ | Reynolds number based on volume |
| $\mathrm{Re}_L$ | Reynolds number based on length |
| $R_n$ | cavitator radius at the section of the cavity origine |
| $S$ | range |
| $S_c$ | surface area of the conical cavitator |
| $S^*$ | maximum range of inertial motion |
| $\bar{S}^*$ | dimensionless range of inertial motion, Eq. (10) |
| $t$ | time |
| $U$ | instantaneous velocity of body |
| $U_i$ | dimensionless initial velocity |
| $U_0$ | initial velocity |
| $U_6$ | velocity at 6 m depth |
| $\bar{U}$ | dimensionless final velocity |
| $\bar{U}^*$ | dimensionless optimal values of final velocity, Eq. (9) |
| $V$ | hull volume |
| $\nu$ | kinematic viscosity of water |
| $V_c$ | volume of the cavitator |
| $\nu_g$ | kinematic viscosity of gas |
| $X$ | total drag of the hull |
| $x$ | dimensionless coordinate |
| $x_{\max}$ | coordinate, corresponding to the maximum cavity radius, Eq. (15) |
| $\alpha, \beta, \omega$ | parameters determined by Eqs. (13), (14) |
| $\alpha_1$ | compressibility correction function, Eq. (19) |
| $\alpha_L$ | correction coefficient for the laminar flow, Eq. (20). |
| $\alpha_T$ | correction coefficient for the turbulent flow, Eq. (21) |
| $\gamma$ | ratio of total drag to pressure drag on cavitator |
| $\theta$ | half angle of the conical cavitator |
| $\Delta D$ | diameter increment of the cavity |
| $\Delta R$ | radius increment of the cavity |
| $\lambda$ | aspect ratio of the cavity |
| $\rho$ | liquid density |
| $\rho_0$ | reference liquid density (at atmospheric pressure, $p_a$) |
| $\rho_g$ | reference gas density (at atmospheric pressure, $p_a$) |
| $\sigma$ | cavitation number |
| $\sigma_f$ | final cavitation number |
| $\sigma^*$ | optimal final cavitation number |
| $\bar{\sigma}^*$ | normalized optimal value of final cavitation number, Eq. (11) |

## Figure Captions List

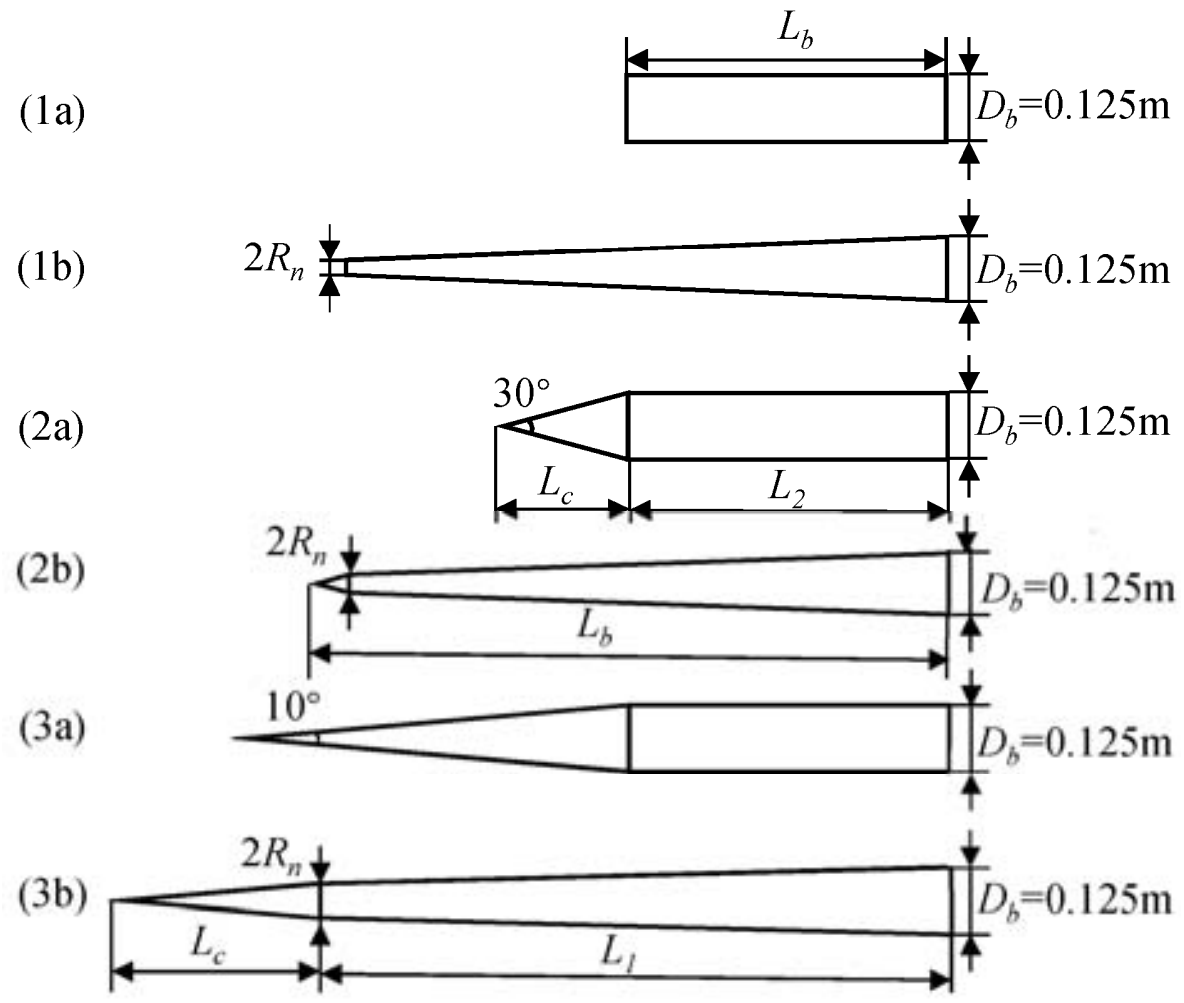

**Fig. 1** Initial (1a, 2a, 3a) and optimized (1b, 2b, 3b) hull shapes

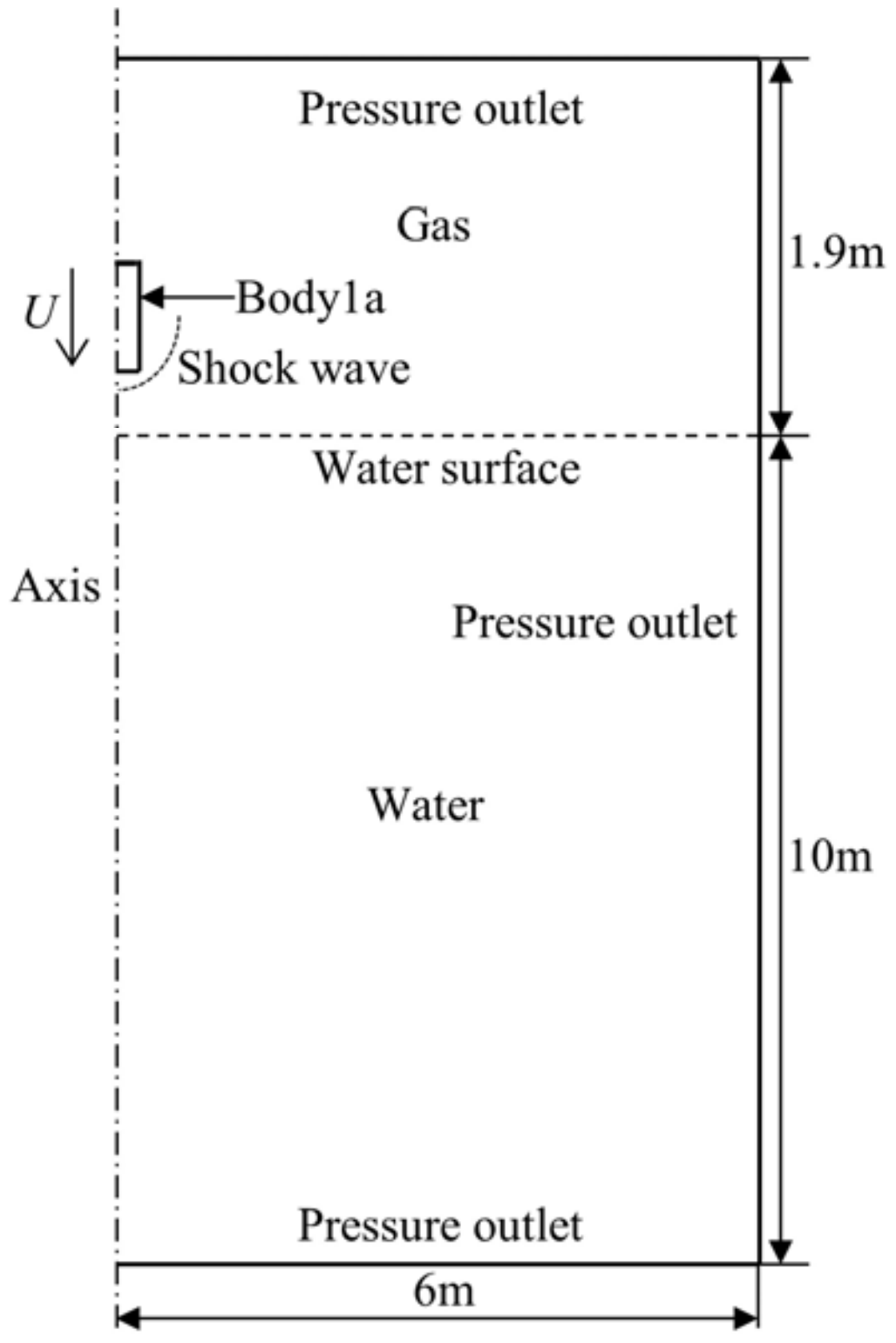

**Fig. 2** Calculation domain and boundary conditions (for initial hull shape 1a with the disc cavitator)

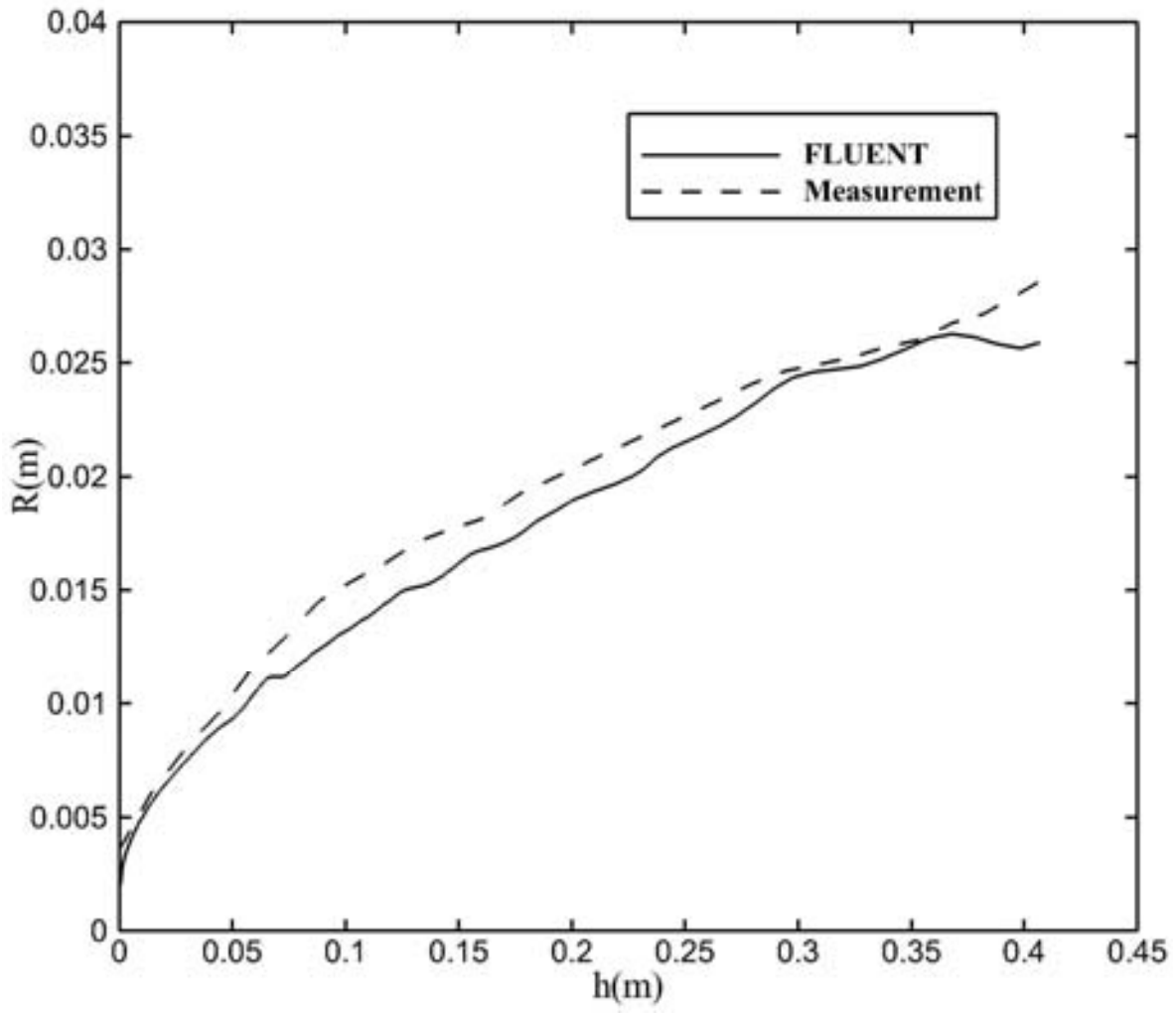

**Fig. 3** Comparison of the calculated and experimental cavity shapes for truncated cone with disc cavitator

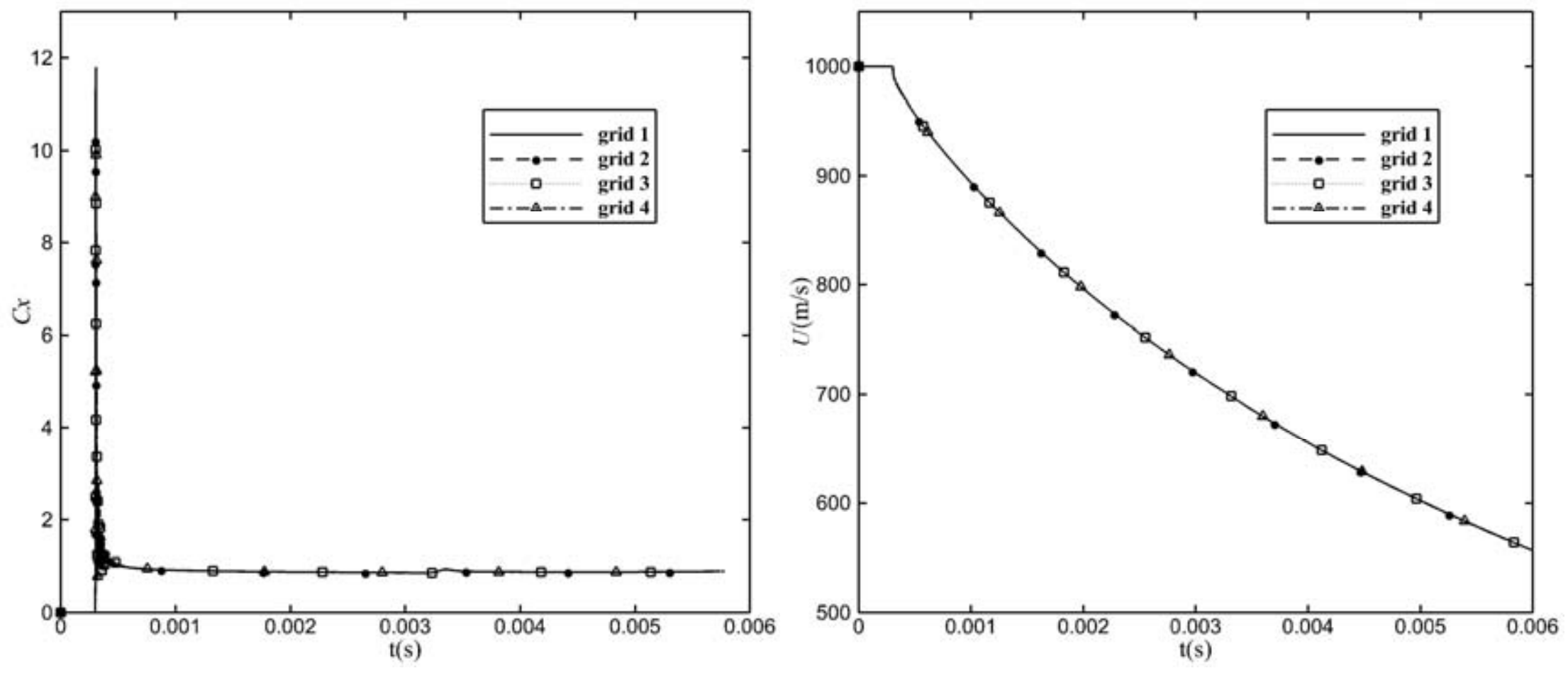

(a) (b)

**Fig. 4** Total drag coefficient $C_x$ (a) and velocity (b) versus time for body 1a and with four different mesh sizes

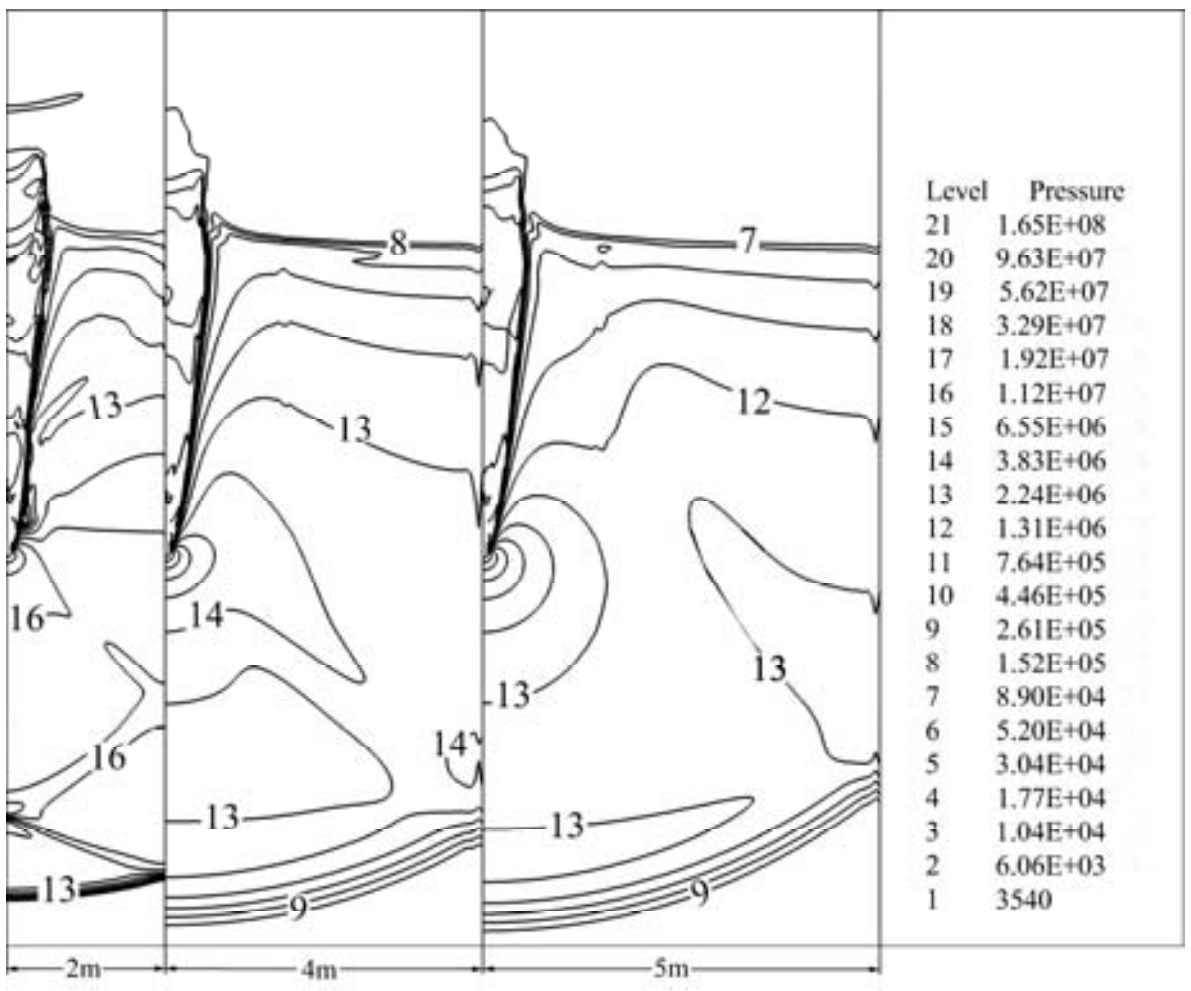

**Fig. 5** Comparison of pressure distribution (in log scale) for different widths of computing domain (body 1a)

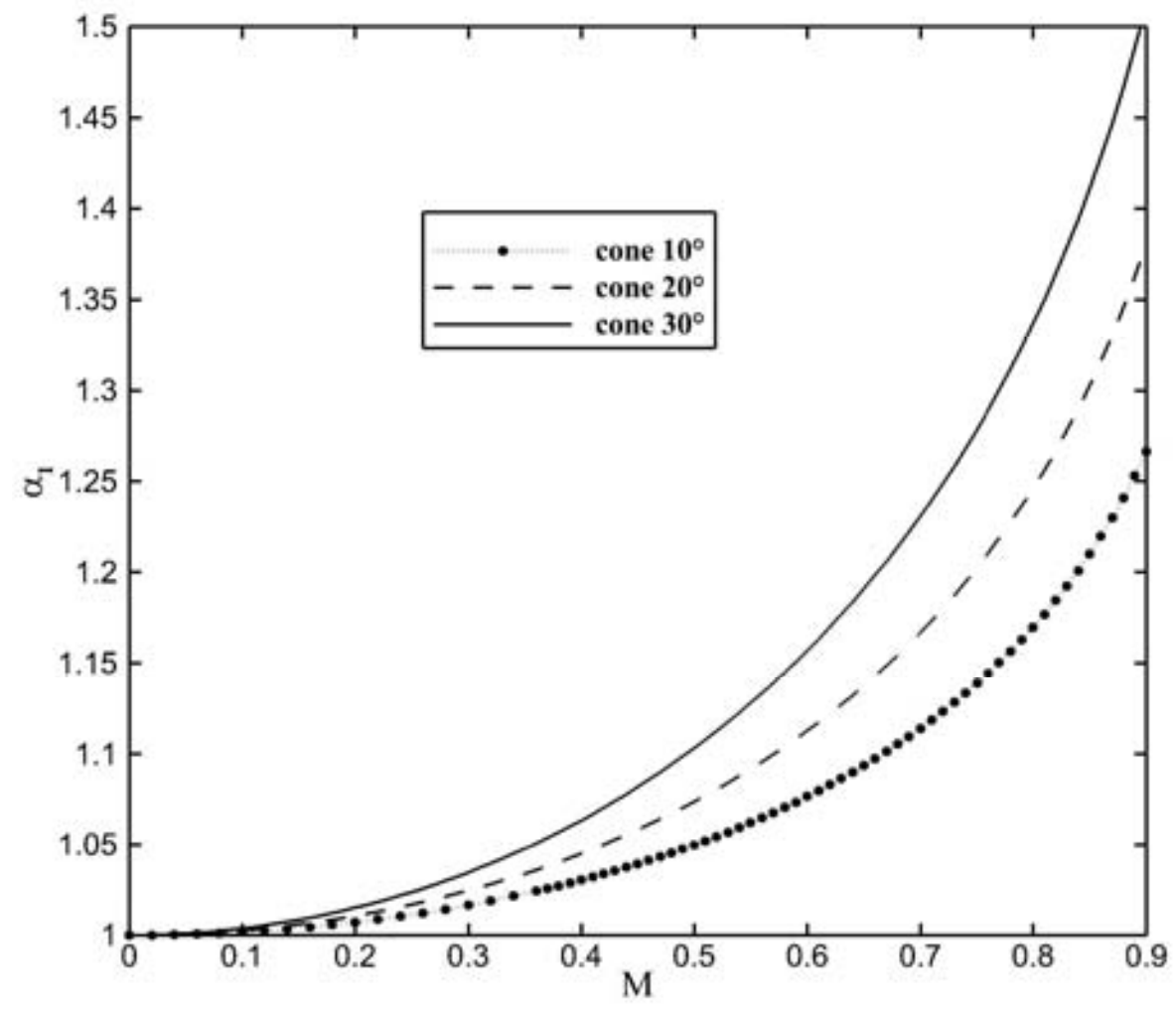

**Fig. 6** Compressibility correction coefficient versus Mach number for different conical cavitators

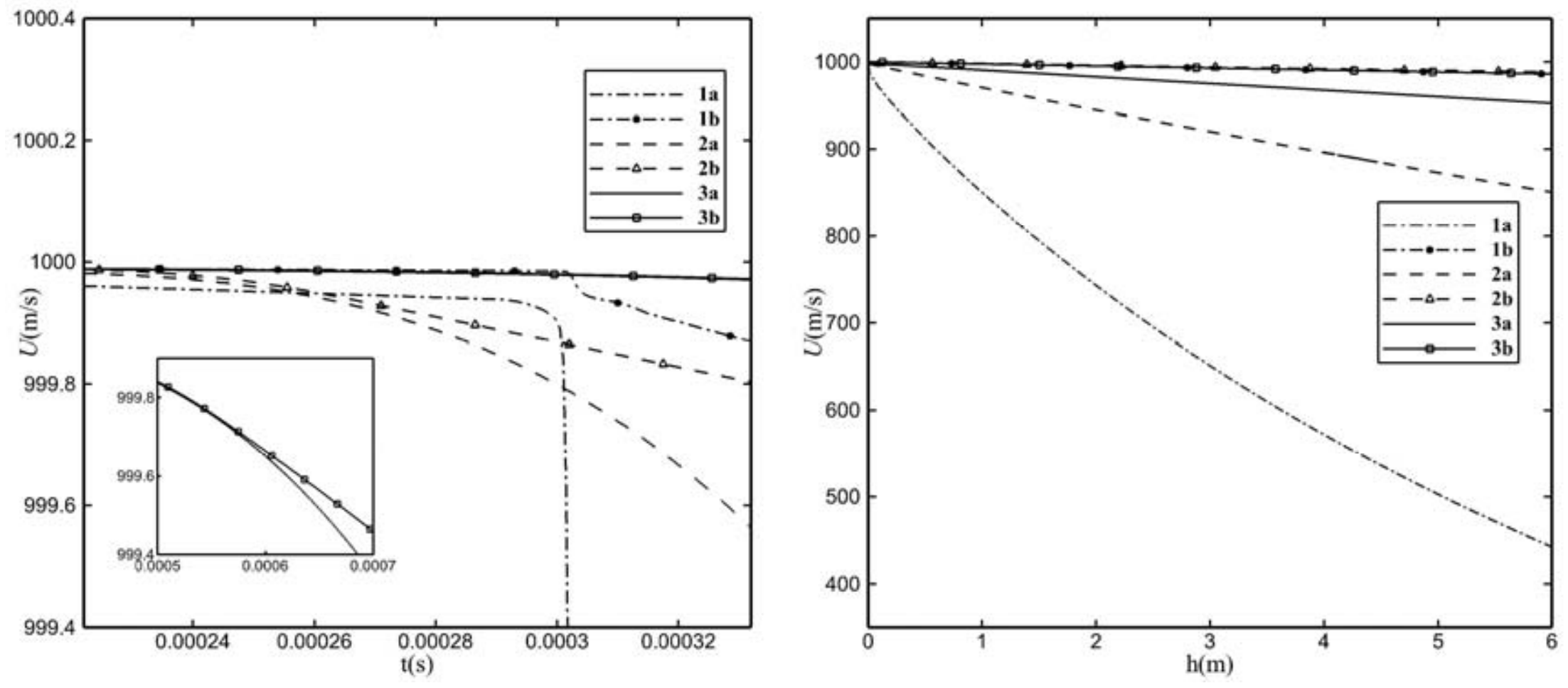

(a) (b)

**Fig. 7** Velocity attenuation versus time (a) and depth (b)

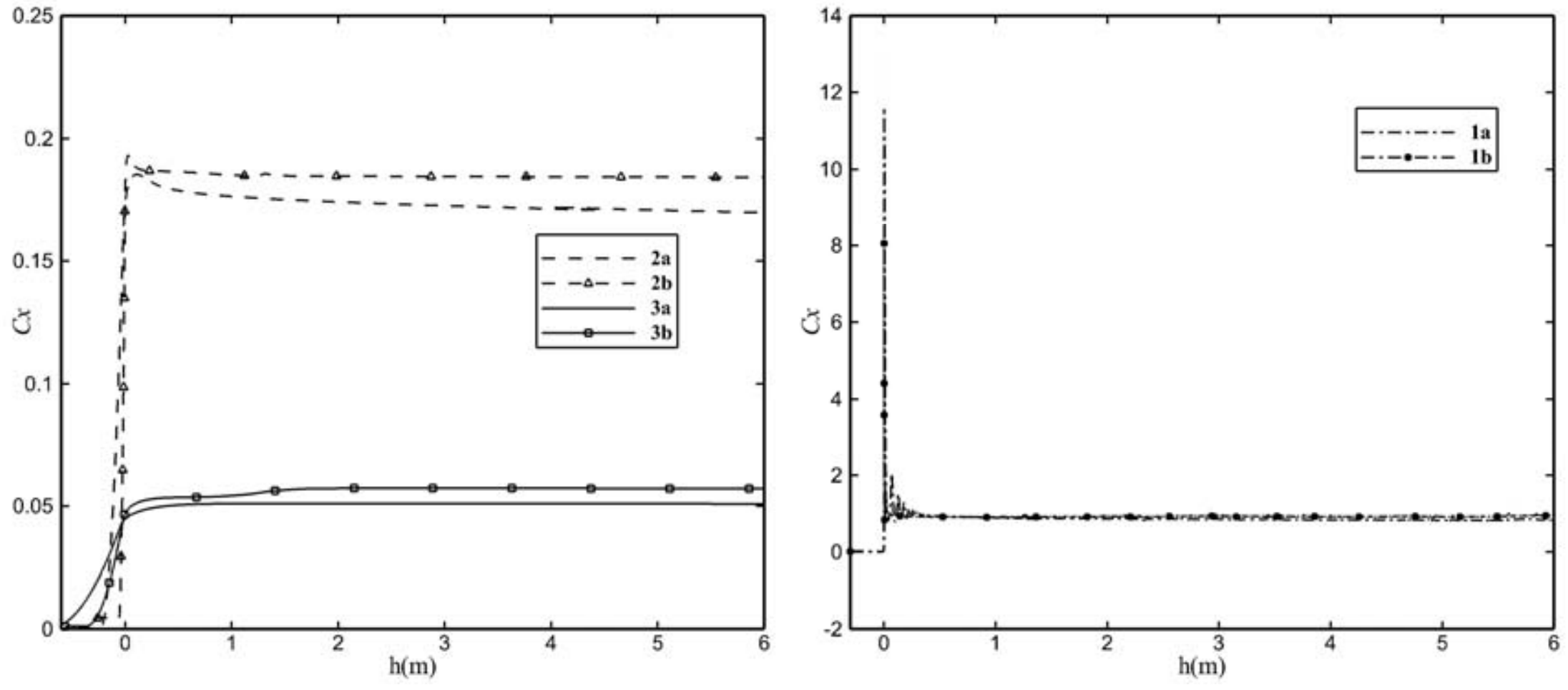

(a) (b)

**Fig. 8** Total drag coefficient $C_x$ versus depth. (a) cone cavitators. (b) disc cavitators

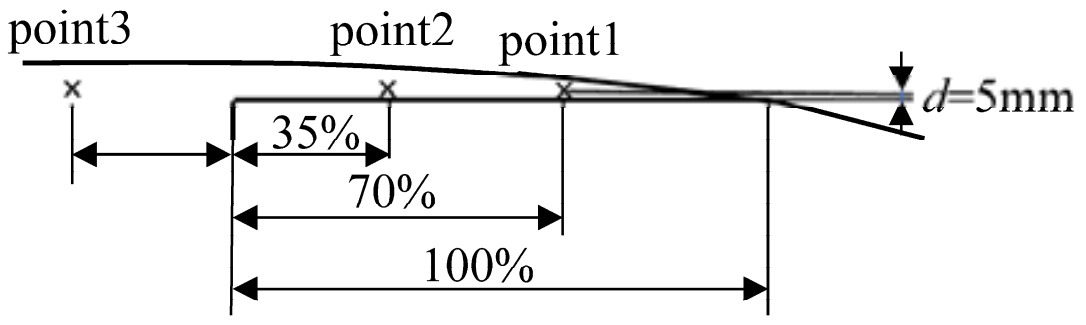

**Fig. 9** Selected points in the cavity

**Table Caption List**

**Table 1** Results of calculations for the initial and optimized hull shapes.

| N | Bodies / Characteristics | Disc 1a | Opt_disc 1b | Cone 30° 2a | Opt_30° 2b | Cone 10° 3a | Opt_10° 3b |
|---|---|---|---|---|---|---|---|
| 1 | Cavitator radius $R_n$, mm | 62.5 | 8<br>7.8 | 62.5 | 16.5<br>17.9 | 62.5 | 32.155<br>32.47 |
| 2 | $L_c$ , mm | 0 | 0 | 233.26 | 61.58 | 714.4 | 367.54 |
| 3 | $L_1$, mm | 0 | 1120<br>1101 | 0 | 1120<br>1102 | 0 | 1185<br>1172 |
| 4 | $L_2$, mm | 600 | 0 | 600 | 0 | 600 | 0 |
| 5 | Volumes $V$ and $V_c$, dm$^3$ | 7.36<br>0 | 5.24<br>0 | 8.32<br>0.95 | 6.13<br>0.018 | 10.29<br>2.92 | 9.02<br>0.40 |
| 6 | $S_c$, cm$^2$ | 0 | 0 | 474.15 | 33.05 | 1408.04 | 372.69 |
| 7 | Speed $U_6$ , m/s | 442.14 | 986.37 | 850.68 | 988.17 | 952.72 | 986.13 |
| 8 | Final speed m/s | 52.0 | 606.6<br>598.3 | 121.3 | 606.6<br>599.4 | 214.3 | 606.6<br>598.1 |
| 9 | $Re_V$, Eq. (22), $Re_L$, Eq. (23) for cavitator | 0<br>0 | 0<br>0 | 8.28E+07<br>1.96E+08 | 2.56E+07<br>6.02E+07 | 1.35E+08<br>6.74E+08 | 7.19E+07<br>3.59E+08 |
| 10 | $C_{xf}$, CFD, theory (laminar and turbulent) | 0<br>0<br>0 | 0<br>0<br>0 | 0.0055<br>0.00041<br>0.0078 | 0.0079<br>0.00075<br>0.0092 | 0.0168<br>0.00067<br>0.0193 | 0.0181<br>0.00093<br>0.0211 |
| 11 | $C_{xp}$ CFD, Eq. (14) | 0.8287<br>(0.82) | 0.895<br>(0.82) | 0.162<br>0.174 | 0.167<br>0.187 | 0.0336<br>0.0367 | 0.0340<br>0.0371 |
| 12 | $C_x$, CFD | **0.8294** | **0.896** | **0.169** | **0.181** | **0.0507** | **0.0565** |
| 13 | $Q_s$ , Eq. (3) | 0.0763 | 0.0025 | 0.0207 | 0.0021 | 0.0068 | 0.0022 |
| 14 | $\gamma$ (CFD) | 1.008 | 1.001 | 1.043 | 1.084 | 1.509 | 1.661 |
| 15 | $\alpha_1$, Eq. (19) | (1.03) | (1.20) | 1.14 | 1.20 | 1.09 | 1.10 |
| 16 | $\alpha_L$ Eq. (20) | 1 | 1 | 1.01 | 1.02 | 1.05 | 1.09 |
| 17 | $\alpha_T$ Eq. (21) | 1 | 1 | 1.04 | 1.05 | 1.53 | 1.57 |
| 18 | Final depth $h_f$ , m | **22.9** | **198.9**<br>**202.8** | **81.3** | **182.3**<br>**187.6** | **198.2** | **132.3**<br>**135.4** |
| 19 | Maximum pressure, atm | **5.64E+04** | **5.51E+04** | **3.07E+03** | **2.94E+03** | **743.78** | **730.11** |
| 20 | $C_v$, CFD | 0.269 | 0.00597 | 0.0505 | 0.00463 | 0.0132 | 0.00424 |

**Table 2** Pressure inside the cavity for different hull shapes at the moments when the bottom of the cavitator achieved the depth of 6 m

| name | | Disc 1a | Opt_disc 1b | Cone 30° 2a | Opt_30° 2b | Cone 10° 3a | Opt_10° 3b |
|---|---|---|---|---|---|---|---|
| $p_c$ (kPa) | point1 | 15.04 | 12.91 | 3.54 | 3.54 | 3.54 | 3.54 |
| | Point2 | 15.32 | 8.29 | 3.54 | 3.54 | 3.54 | 3.54 |
| | Point3 | 39.03 | 37.40 | 5.89 | 34.63 | 22.07 | 41.72 |